\documentclass[aps,superscriptaddress,twocolumn,showpacs,preprintnumbers,amsmath,amssymb,showkeys]{revtex4}
\usepackage[cp1251]{inputenc}
\usepackage{amssymb,amsmath}
\usepackage[mathscr]{eucal}
\usepackage{graphicx}
\usepackage{longtable}
\usepackage[dvips]{hyperref} 
\begin{document}

\title{Quantum-mechanical equation for spectroscopic transitions in ordered ferroelectric and ferromagnetic chains }

\author{Dmitry Yearchuck (a), Yauhen Yerchak (b), Alla Dovlatova (c)\\
\textit{(a) - Minsk State Higher Aviation College, Uborevich Str., 77, Minsk, 220096, RB; yearchuck@gmail.com, \\ (b) - Belarusian State University, Nezavisimosti Ave., 4, Minsk, 220030, RB; \\ (c) - M.V.Lomonosov Moscow State University, Moscow, 119899}}
	
\date{\today}
             
\begin{abstract} Transition operator method is proposed for description of the dynamics of spectroscopic transitions.
Quantum-mechanical analogue of Landau-Lifshitz equation has been derived for the system representing 
itself the periodical ferroelectrically (ferromagnetically) ordered chain of 
$N$ equivalent elements, interacting with external oscillating electromagnetic 
field. Landau-Lifshitz equation was represented in Lorentz invariant form by using Hilbert space over the ring of quaternions. It has been  shown, that spin vector can be considered to be quaternion vector of the state of the system studied. From comparison with experiment for the first time from pure optical measurements the value of spin $S = 1/2$ for optically active centers - spin-Peierls solitons in carbon chains - has been obtained. The ratio of imagine  to real   components of complex charge is evaluated for given centers to be $\frac{g}{e} \approx (1.1 - 1.3)10^{2}$.
\end{abstract}

                             
\maketitle
\section{Introduction}
Spectroscopic description of the systems, interacting with electromagnetic (EM) field including the description of transitional effects,
for instance, Rabi-oscillations, free induction, 
spin echo in magnetic resonance spectroscopy and
corresponding optical analogues in optical spectroscopy
is achieved within the framework of gyroscopic model. Mathematical base for gyroscopic model is Landau-Lifshitz (L-L) 
equation. L-L equation was postulated by Landau and Lifshitz for macroscopic 
classical description of the motion of magnetization vector in ferromagnets 
in 1935 \cite{Landau}. L-L equation is:
\begin{equation}
\label{eq1}
\frac{d\vec {S}}{dt}=[\gamma _{_H} \vec {H}\times \vec {S}],
\end{equation}
where $\vec {S}$ is magnetic moment, $\vec {H}$ is effective magnetic field, 
$\gamma _{_H} $ is gyromagnetic ratio. 
Clasical gyroscopic
model for transition dynamics and for description
of transitional effects was introduced formally \cite{Abragam} by
F.Bloch, who has added the phenomenological relaxation term to L-L equation.
Then L-L equation and Bloch equations, based on it, were substantiated 
quantum-mechanically in magnetic resonance theory, but  the only for  
describtion of the motion of magnetic moment vector in external magnetic 
field \cite{Abragam}, \cite{Slichter}. Quantum-mechanical Bloch equations were in fact postulated for the description of dynamics
of magnetic resonance transitions. The development of gyroscopic
model for the characterization of optical systems, interacting with EM-field was done  on the base of density operator formalism, that is quantum statistically \cite {Apanasevich}, \cite {Scully}. 

The optical analogue of L-L equation is:
\begin{equation}
\label{eq2}
\frac{d\vec {P}}{dt}=[\vec {P}\times \gamma _{_E} \vec {E}],
\end{equation}
where $\gamma _{_E}$ was called formally gyroelectric ratio by using the only its mathematical analogy with gyromagnetic ratio without however corresponding physical meaning,  $\vec {P}$ (Bloch vector), $\vec {E}$ are some axial vectors, with components representing themselves various physical quantities. So, for instance, in \cite{Macomber} is emphasized that the only $P_{x}$, 
$P_{y}$ and $E_{x}$, $E_{y}$ components of vectors $\vec {P}$, $\vec {E}$, correspondingly, 
characterize the genuine electromagnetic properties of the system, at the same time the components $P_{z}$ and $E_{z}$ cannot be referred to electromagnetic characteristics. In other words physical meaning of Bloch vector and $\vec {E}$
 seems in existing optical transition theory to be unclear. In other words, the vectors $\vec {P}$, $\vec {E}$ entering
optical Bloch equations are in fact some mathematical
abstractions. Really, like
to magnetic moment and magnetic field vectors in magnetic resonance Bloch equations, both the vectors $\vec {P}$, $\vec {E}$ have, according to existing description, necessary axial symmetry, however in artificial way. For instance, $\vec {P}$  has two components $P_{x}$, $P_{y}$ of polar electric dipole moment vector, that is, both the components remain without any physical meaning change. At the same time the third component of $\vec {P}$ has entirely different physical meaning and it is considered to be the population difference between the energy levels, see for example \cite {Apanasevich}, \cite{Scully}. Mathematically the objects, which are like to Bloch vector, can, naturally, exist. However, even mathematically, it will mean, that the spaces, which produce the set of Bloch vectors ${\vec {P}}$ and the set ${\vec {E}}$ over the field of real numbers, is the only affine space and it cannot be metrizable. In other words, the components of $\vec {P}$, $\vec {E}$ in existing interpretation cannot represent the physical quantity to be a measurable single whole.  
 Physically the model, in which the part of components of characteristic vector possess by different symmetry (polar in given case) in comparison with the final symmetry (axial) of vector itself seems to be also incorrect. Given conclusion is especially evident, if to take into account,
that the mechanism of  spectroscopic transitions' dynamics itself has
to be independent on the nature  of the centers, interacting with free oscillating electromagnetic field (EM-field), that is, whether they possess by magnetic or electric dipole moments, owing to independence of magnetic and electric components of 
EM-field. Given conclusion is in complete agreement with experiment, since dynamics of the spectroscopic transitions both in otical and magnetic resonance areas can be described mathematically by the same phenomenological Bloch equations.
Let us remember, that the analogue of Bloch vector in magnetic resonance spectroscopy is total magnetic moment, that is pure axial vector with peer all the three components. 

Note, that most general at present description of spectroscopic transitions' dynamics by means of density operator formalism has a substantial disadvantage.
Really, it seems to be understandable, that in some experimental cases, in particular, for local single 
centers, which can be used in nanotechnology, for
instance, for elements of binary logic, quantum informatics, quantum  computing with very fast rates of processes of information treatment, realized on semiconductor QDs, Josephson cubits, spin-qubits and so on,  statistical density operator formalism
can be not applicable for description of transitions' dynamics at all.  The main reason is the following. If the state of quantum system is mixed state, that seems to be necessary for functioning of nanotechnology elements, then the event frequency of realization of definite pure state in the superposition can be not corresponding to its probability. In other words, the well known law of probabilities' theory - the law of large numbers has to be taken into account. Moreover, by measurements with short times, which are comparable with the residence time in definite pure state iof the superposition, the amplitude of given state cannot be determined by its probability for any quantum system.

New approach to a description of spectroscopic transitions' dynamics, which allows to overcome given difficulties, was proposed for the first time in \cite{Yearchuck_Doklady}. Given approach is based on transition operator method. 
Operator equation, describing the optical transitions' dynamics (for the example of simple 1D-model of quantum system), has  been obtained by using of stated method \cite{Yearchuck_Doklady}. It has been shown, that given equation is operator equivalent to Landau-Lifshitz (L-L) equation \cite{Landau} in its 
difference-differential form, which takes usual differential form in continuum limit.

The aim of the work presented is description of transition operator method in more details and development of the model, proposed in \cite{Yearchuck_Doklady}.  
\section{Spectroscopic transitions' model for 1D-chain of exchange interacting centers}

Let us consider the system representing 
itself the periodical ferroelectrically (ferromagnetically) ordered chain of 
$N$ equivalent elements, interacting with external oscillating electromagnetic 
field (EM-field). It is assumed, that the interaction between elements (elementary 
units) of the chain can be described by the Hamiltonian of quantum XYZ 
Heisenberg model in the case of a chain of magnetic dipoles and by 
corresponding optical analogue of given Heisenberg model in the case of a 
chain of electric dipoles. We will consider for the simplicity the case of 
isotropic exchange. Each elementary unit of the chain will be considered to be 
two-level quantum system like to one-electron atom. Then Hamiltonians for the 
chain of electrical dipole moments and for the chain of magnetic dipole 
moments will be mathematically equivalent. The chain for distinctness of 
electrical dipole moments can be described by the following Hamiltonian

\begin{equation}
\label{eq4}
{\mathcal{\hat H}} = {\mathcal{\hat H}^0} + \mathcal{\hat H}^{C F} + {\mathcal{\hat H}^J}, 
\end{equation} 
where ${\mathcal{\hat H}^0}$ is chain Hamiltonian in the absence of the interaction with external EM-field and between structural elementary units of the chain themselves.
We suggest, that given structural elementary units are optical (magnetic) centers with one electric (magnetic) dipole pro center. ${\mathcal{\hat H}^{C F}}$ is Hamiltonian, describing the interaction between classical external EM-field and atomic chain, ${\mathcal{\hat H}^J}$ is Hamiltonian of exchange interaction between above indicated  centers. 

To obtain the equation, describing the transition dynamics, we will use the transition operator method. Given method was used in atomic spectroscopy, see, for instance, \cite{Scully}. The essence of the method, adapded for our task, is the following. Let $\left\{ {\left|m_l \right\rangle}\right\}$, $l = \overline{1,n}$, $m = \alpha, \beta$ be full set of energy states of $l-\textit{th}$ optical (magnetic) center of the chain in the absence of the interaction both between the centers themselves and with EM field. Here $\left| {\alpha _l} \right\rangle $ is lower state and $\left|{\beta _l } \right\rangle $ is upper state. They are eigenstates of $l-\textit{th}$ item in ${\mathcal{\hat H}^0}$, $l = \overline{1,n}$. It means, that ${\mathcal{\hat H}^0}$ is
\begin{equation}
\label{eq5}
\mathcal{\hat H}^0 = \sum\limits_{v=1}^N \sum\limits_m E_{mv}{\left|m_v \right\rangle} {\left\langle m_v \right|}.
\end{equation}
Here $m = \alpha, \beta$, $E_{mv}$ are eigenvalues of ${\mathcal{\hat H}^0}$, which correspond to the states ${\left|m_v \right\rangle}$ of $v-\textit{th}$ chain center. 
Let set up in correspondence to the states ${\left|j_v \right\rangle}$, $v = \overline{1,n}$ the operators 
\begin{equation}
\label{eq6}
{\hat\sigma_v}^{jm} \equiv {\left|j_v \right\rangle} {\left\langle m_v \right|}, 
\end{equation}
where $j = \alpha, \beta, m = \alpha, \beta $.
Then, it is evident, that operators $\hat {\sigma }_v^+ = \left| {\beta _v } \right\rangle 
\left\langle {\alpha _v } \right|$ are the operators, which transform eigenstates $\left| {\alpha_v } \right\rangle$ into $\left| {\beta_v }\right\rangle$ and operators $\hat {\sigma }_v^- = \left| {\alpha_v } \right\rangle \left\langle {\beta _v} \right|$ are the operators, transforming eigenstates $\left| {\beta_v } \right\rangle$ into $\left| {\alpha_v }\right\rangle$ for $\forall v$, $v = \overline{1,n}$. Let define also the operators $\hat {\sigma }_v^z = \left| {\beta _v } \right\rangle \left\langle {\beta _v 
} \right| - \left| {\alpha _v } \right\rangle \left\langle 
{\alpha _v } \right|$. The operators $\hat {\sigma }_v^z$ transform for $\forall v$, $v = \overline{1,n}$ the sum of states $\left| {\beta_v } \right\rangle$ + $\left| {\alpha_v } \right\rangle$ into their difference and the reverse is also true. The above defined set of spectroscopic transition operators can be completed by unit operators $\hat {\sigma }_v^ E$, $v = \overline{1,n}$, which, owing to completeness of state sets can be represented by the following relationship
\begin{equation}
\label{eq7}
\hat {\sigma }_v^E = \sum\limits_m \left| {m_v } \right\rangle \left\langle {m_v} \right|,
\end{equation}
 $m = \alpha, \beta $, 
and by zeroth operators $\hat {\sigma}_v^ 0$, $v = \overline{1,n}$, which can be defined in the following way
\begin{equation}
\label{eq8}
\hat {\sigma }_v^0 = \hat {\sigma }_v^j - \hat {\sigma }_v^j,
\end{equation}
where $j = -, +, z, E$.
It is easy to show, that at fixed $v$ the set of transition operators is closed relatively the algebraic operations of commutation and anticommutation (that is relatively taking of commutator and anticommutator) and relatively the operation of Hermitian conjugation. For instance, the relationships for commutation rules are
\begin{equation}
\label{eq9a}
[\hat {\sigma}_v^{lm}, \hat {\sigma}_v^{pq}] = \hat {\sigma }_v^{lq} \delta_{mp} - \hat {\sigma }_v^{pm}\delta_{ql}, 
\end{equation}
that is, 
\begin{equation}
\label{eq9b}
[\hat {\sigma}_v^{-}, \hat {\sigma}_v^{z}] = 2 \hat {\sigma }_v^{-}, 
[\hat {\sigma}_v^z, \hat {\sigma }_v^{+}] = 2 \hat {\sigma }_v^+, 
[\hat {\sigma}_v^{+}, \hat {\sigma }_v^{-}] = \hat {\sigma }_v^z. 
\end{equation}
 So,  at fixed $v$ the set of transition operators $\left\{{\hat\sigma_v}^{jm}\right\}$ produces algebra. In particular, the set 
of $\left\{\hat {\sigma }_v^m\right\}$ operators, where m is z, +, -, 0, E, produces algebra, 
which is isomorphic to $S = 1/2$ Pauli matrix algebra, completed by unit and zeroth matrices, that is, 
mappings
\begin{equation}
\label{eq9}
 f_v :\,\hat {\sigma }_v^m \to \sigma _P^m, m = z, +, -, 0, E 
\end{equation}
 realize 
isomorphism. Here $v$ is a number of chain unit, $\sigma _P^m $ 
is extended set of Pauli matrices for the spin of $1/2$.  
unit.

 The structure of transition operators indicates on mathematical advantages of transition operators' method in comparison with the method of density matrix. Really, the method of density matrix gives the relationship for probabilities of the states, in spectroscopic transition operators' method we are dealing immediately with the states. 

Let us define the vector operator: 
\begin{equation}
\label{eq10a}
\hat {\vec {\sigma }}_v =\hat {\sigma }_v^- \,\vec {e}_+ +\hat {\sigma 
}_v^+ \,\vec {e}_- +\hat {\sigma }_v^z \,\vec {e}_z .
\end{equation}
Consequently, taking into account (\ref{eq9}), from 
physical point of view $\hat {\vec {\sigma }}_v$ represents itself some 
vector operator, which is proportional to vector operator of the spin of $v-\textit{th}$ chain 
unit.

 It is evident, that the set of vector operators $\left\{\hat {\vec {\sigma }}_v\right\}$ produce linear space over field of complex numbers, which 
can be called transition space. It is 3-dimensional (in the case of two-level systems), that is 3 matter operator 
equations of the motion for components of $\hat {\vec {\sigma }}_v$ is necessary for correct 
description of optical transitions in semiclassical approach. Typical inaccuracy in many of the quantum 
optics calculations of transition dynamics is connected with obliteration of correct dimensionality of transition space.
Further, isomorphism of Pauli matrix algebra and spectroscopic transition operators' algebra allows to change the spin variables in known Heisenberg exchange Hamiltonian by components of vector $\hat {\vec {\sigma }}_v$.

We will use further the rotating wave approximation \cite{Scully}. Then the chain for distinctness of 
electrical dipole moments can be described by the following Hamiltonian:
\begin{equation}
\label{eq10}
\begin{split}
\raisetag{40pt}
&\mathcal{\hat H} = \frac{\hbar \omega _0}{2}\sum\limits_{n = 1}^N {\hat {\sigma}_n^z } - p_{_{E}}^{\alpha \beta }\sum\limits_{n = 1}^N {E_1^n (\hat {\sigma} _n^ + e^{ - i\omega t} + \hat {\sigma} _n^- e^{i\omega t} )}\\
& \ \  + \sum\limits_{n = 1}^N [J_{_{E}} (\hat {\sigma} _n^ + \hat {\sigma} _{n + 1}^- + \hat {\sigma} _n^ -  \hat {\sigma} _{n + 1}^ + + \frac{1}{2}\hat {\sigma} _n^z \hat {\sigma} _{n + 1}^z ) + H.c.].
\end{split}
\end{equation}
It is suggested 
in the model, that $\left| {\alpha _n } \right\rangle $ and $\left| {\beta _n } \right\rangle $ are eigenstates, producing the full set for each of $N$ elements. It is evident, that given assumption can be realized strictly the 
only by the absence of the interaction between the elements. At the same 
time proposed model will rather well describe the real case, if the 
interaction energy of adjacent elements is much less of the energy of the 
splitting $\hbar \omega _0 =\mathcal{E}_\beta -\mathcal{E}_\alpha$ between the energy levels, 
corresponding to the states $\left|\alpha_n\right\rangle$ and $\left|\beta_n\right\rangle$. This case includes in fact all known 
experimental situations. Further, $p_{_E}^{\alpha \beta }$ is matrix element 
of dipole transitions between the states $\left| {\alpha _n } \right\rangle $ and $\left| {\beta _n } \right\rangle $. Matrix elements along with energy difference of given states $\mathcal{E}_\beta -\mathcal{E}_\alpha$ are suggested to be 
independent on $n$, $E_1^n $ is amplitude of electric component of external
EM-field on the $n$\textit{-th} element site, $\hbar$ is Planck's constant, $J_{E}$ is optical analogue of the 
exchange interaction constant. Here, in correspondence with the suggestion, 
$J_{_E} =J_{_E}^x =J_{_E}^y =J_{_E}^z $. In the case of the 
chain of magnetic dipole moments  electrical characteristics of the systems studied in Hamiltonian (\ref{eq10}) are  replaced by corresponding magnetic
characteristics
\begin{equation}
\label{eq3}
\begin{split}
\raisetag{60pt}
&E_1^n \rightarrow H_1^n,\\
&J_{_E} \rightarrow J_{_H},\\
&p_{_E}^{\alpha \beta} \rightarrow p_{_H}^{\alpha \beta},\\
&\omega_{0} \rightarrow \frac{1}{\hbar}g_{_H} \beta_{_H} H_{0} = \gamma_{_H} H_{0}, 
\end{split}
\end{equation}
where $ H_1^n$ is amplitude of magnetic component of external
EM-field (microwave field) on the 
$n$\textit{-th} element site, $J_{_H}$ is  the constant of magnetic spin exchange interaction,  
$p_{_H}^{\alpha \beta}$ is  matrix element of  magnetic resonance dipole transitions, $H_0$ is external static magnetic field, $\beta_{_H}$ is Bohr magneton, $g_{_H}$ is $g$-tensor, which is assumed for the simplicity to be isotropic. 

The first term in Hamiltonian (\ref{eq10}) characterizes the total energy of all chain elements in the absence of external field and in the absence of interaction 
between chain elements. The second item characterizes an interaction of a 
chain with an external oscillating EM-field in dipole 
approximation. Matrix elements of full dipole up and down transitions between couples of the states ($\left|{\alpha_n} \right\rangle $, $\left|{\beta_n} \right\rangle$) and ($\left|{\beta_n} \right\rangle$, $\left|{\alpha_n} \right\rangle $), respectively, are assumed to be equal, that is, spontaneous emission is not taken into consideration. The third item is, in essence, Hamiltonian of quantum Heisenberg \textit{XXX}-model in the case of magnetic 
version and its electrical analogue in electric version of the model 
proposed. 
The equation of the motion for $\hat {\vec {\sigma }}_k$ is:
\begin{equation}
\label{eq12}
i\hbar \,\frac{{\partial \hat {\vec {\sigma} }_k}}{{\partial t}} = [\hat {\sigma} _k^ -  ,\mathcal{\hat H} ]\, \vec e_ +   + [\hat {\sigma} _k^ +  ,\mathcal{\hat H} ]\, \vec e_ -   + [\hat {\sigma} _k^z ,\mathcal{\hat H} ]\, \vec e_z. 
\end{equation}
We will consider the case of homogeneous excitation of the chain, that is $E_1^n $ is independent on unit number $n$ ($E_1^n \equiv E_1)$. Then, using the commutation relations
$[\hat {\sigma }_k^+ ,\hat {\sigma }_n^- ]=\delta _{kn} \hat {\sigma }_n^z 
,\,[\hat {\sigma }_n^- ,\hat {\sigma }_k^z ]=2\delta _{kn} \hat {\sigma 
}_n^- ,\,[\hat {\sigma }_k^z ,\hat {\sigma }_n^+ ]=2\delta _{kn} \hat 
{\sigma }_n^+$,
we obtain
\begin{subequations}
\label{eq13}
\begin{gather}
\frac{{\partial \hat\sigma _k^z }}{{\partial t}} = 2i\Omega _{_E} \left( {e^{ - i\omega t} \hat\sigma _k^ + - e^{i\omega t} \hat\sigma _k^ - } \right)\\ 
+\frac{{2iJ_{_E} }}{\hbar }\left(\left\{ {\hat\sigma _k^ - } \right.,\left. {(\hat\sigma _{k + 1}^ + + \hat\sigma _{k - 1}^ + )} \right\} - 
\left\{ {\hat\sigma _k^ + } \right.,\left. {(\hat\sigma _{k + 1}^ - + \hat\sigma _{k - 1}^ - )} \right\}\right),\nonumber
\\
\frac{{\partial \hat\sigma _k^ +  }}{{\partial t}} = i\omega _0 \hat\sigma _k^ +   + i\Omega _{_E}  e^{i\omega t} \hat\sigma _k^z 
\\
+\frac{{iJ_{_E} }}{\hbar } \left(\left\{ {\hat\sigma _k^ +  } \right.,\left. {(\hat\sigma _{k + 1}^z  + \hat\sigma _{k - 1}^z )} \right\} - \left\{ {\hat\sigma _k^z } \right.,\left. {(\hat\sigma _{k + 1}^ +   + \hat\sigma _{k - 1}^ +  )} \right\}\right),\nonumber
\\
\frac{{\partial \hat\sigma _k^-}}{{\partial t}} =  - i\omega _0 \hat\sigma _k^ -   - i\Omega _{_E} e^{ - i\omega t} \hat\sigma _k^z  
\\  
-\frac{{iJ_{_E} }}{\hbar }   \left(\left\{ {\hat\sigma _k^ -  } \right.,\left. {(\hat\sigma _{k + 1}^z  + \hat\sigma _{k - 1}^z )} \right\} + \left\{ {\hat\sigma _k^z } \right.,\left. {(\hat\sigma _{k + 1}^ - + \hat\sigma _{k - 1}^ - )} \right\}\right),\nonumber
\end{gather}
\end{subequations}
where expressions in braces $\left\{ {\,,\,} \right\}$ are anticommutators. Here 
$\Omega _{E}$ and $\gamma _{_E} $ are Rabi frequency and gyroelectric ratio respectively. 
They are determined, correspondingly, by relationships 
\begin{equation}
\label{eq11a}
\Omega _{_E} =\frac{E_1 p_{_E}^{\alpha \beta } }{\hbar }=\gamma _{_E} E_1 ,
\quad
\gamma _{_E} =\frac{p_{_E}^{\alpha \beta } }{\hbar },
 \end{equation}
which are
 replaced by relationships
\begin{equation}
\label{eq11b}
 \Omega _{_H} = \frac{H_1 p_{_H}^{\alpha \beta }}{\hbar}=\gamma_{_H} H_1,
\quad
\gamma_{_H} = \frac{p_{_H}^{\alpha \beta }}{\hbar}
 \end{equation}
in the case 
of the chain of magnetic dipole moments. The equations (\ref{eq13}) can be represented in compact vector 
form, at that the most simple expression is obtained by using of the basis 
$\vec {e}_+ =(\vec {e}_x +i\vec {e}_y ), \,\vec {e}_- =(\vec {e}_x -i\vec{e}_y ), \vec {e}_z $. So, we have
\begin{equation}
\label{eq14}
\frac{{\partial \hat {\vec {\sigma}} _k }}{{\partial t}} = \left[ {\hat {\vec {\sigma}} _k \times \hat {\vec {\mathfrak{G}}}_{k - 1,k + 1} } \right],
\end{equation}
where $\hat {\vec {\sigma }}_k$ is given by (\ref{eq10a}), but with the components in matrix representation of operators,
corresponding to new basis, $k = \overline {2,N-1} $, and vector operator 
$\hat {\vec {\mathfrak{G}}}_{k - 1,k + 1}  = \hat {\mathfrak{G}}_{k - 1,k + 1}^-  \vec e_ +   + \hat {\mathfrak{G}}_{k - 1,k + 1}^ +  \vec e_ -   + \hat {\mathfrak{G}}_{k - 1,k + 1}^z \vec e_z$, where its components are
\begin{equation}
\label{eq15}
\begin{split}
\raisetag{60pt}
&\hat {\mathfrak{G}}_{k - 1,k + 1}^-  = \Omega _{_E} e^{-i\omega t} - \frac{{2J_{_E} }}{\hbar }(\hat\sigma _{k + 1}^- + \hat\sigma _{k - 1}^-),  \\
&\hat {\mathfrak{G}}_{k - 1,k + 1}^+ = \Omega _{_E} e^{i\omega t} - \frac{{2J_{_E} }}{\hbar }(\hat\sigma _{k + 1}^+ + \hat\sigma _{k - 1}^ + ),  \\
&\hat {\mathfrak{G}}_{k - 1,k + 1}^z = - \omega_0 - \frac{{2J_{_E} }}{\hbar }(\hat\sigma _{k + 1}^z + \hat\sigma _{k - 1}^z ).
\end{split}
\end{equation}
It should be noted, that vector product of vector operators in (\ref{eq14}) is 
calculated in the correspondence with  expression
\begin{equation}
\label{eq16}
 \left[ {\hat {\vec {\sigma}} _k \times \hat {\vec {\mathfrak{G}}}_{k - 1,k + 1} } \right] = \frac{1}{2}\left| {\begin{array}{*{20}c}
 {\vec e_- \times \vec e_z} & {\hat{\sigma}_k^-} & {\hat {\mathfrak{G}}_{\,\,k - 1,k + 1}^-} \\
 {\vec e_z \times \vec e_+} & {\hat{\sigma}_k^+} & {\hat {\mathfrak{G}}_{\,\,k - 1,k + 1}^+} \\
 {\vec e_+ \times \vec e_-} & {\hat{\sigma}_k^z} & {\hat {\mathfrak{G}}_{\,\,k - 1,k + 1}^z} \\
\end{array}} \right|.
\end{equation}
Given expression is similar to an expression usually used in vector analysis, however, anticommutators of corresponding 
components instead their products are taken by its calculation. In fact it is definition for vector product in 
the case of operator vectors.
Given definition seems to be natural generalization of usual vector product, since the only in given case the result will 
 be independent on a sequence of components of both the vectors in their 
products like to that one for usual vectors. Naturally, the expressions like 
to (\ref{eq16}) can be used for calculation of vector product of common 
vectors, which is provided with coefficient $\frac{1}{2}$. 

Taking into account the physical sense of vector operators $\hat 
{\vec {\sigma }}_k$ we conclude, that (\ref{eq14}) represent themselves 
required quantum-mechanical difference-differential equations (the time is 
varied continuously, the coordinates are varied discretely) for the 
description of the dynamics of the spectroscopic transitions (in the frames 
of the model proposed). It follows from here in view of isomorphism of algebras of 
operators $\hat {\vec {\sigma }}_k $ and components of the spin,  
that the (\ref{eq14}) is equivalent to L-L equation (in its 
difference-differential form). 

Therefore, we have proved, on the one hand, the possibility 
to use L-L equation for the description of the dynamics of spectroscopic 
transitions and for the description of transitional effects from unified theory positions for both cases, that in, in optical and magnetic resonance spectroscopy areas. On the other hand, we have proved (in the frames 
of the model proposed), that the quantum equation 
 for the 
description of the dynamics of the spectroscopic transitions coincides with the equation for spin moment motion. Given result was used always in its implicit form  by spectroscopic transition dynamics studies in magnetic resonance, however without its substantiation, see, for instance \cite{Slichter}. 
 
To obtain the continuous approximation of (\ref{eq14}) for coordinate variables, 
we have to suggest that the length of electromagnetic wave $\lambda $ 
satisfies the relationship $\lambda >> a$, where $a$ is 1D-chain-lattice constant. It is evident, that in optical frequency region the relationship $\lambda >> a$ is always takes place. Then the 
continuous limit will be  realized, if to substitute all the operators, which 
depend on discrete variable $k$, for the operators, depending on continuous 
variable $z$, that is,
\begin{equation}
\label{eq16a}
 \hat{\sigma }_k^\pm \to \hat{\sigma }^\pm 
(z), \hat{\sigma }_k^z \to \hat{\sigma }^z(z). 
\end{equation}
Thus, we obtain, 
taking also into account the relationships 
\begin{equation}
\label{eq16b}
\hat{\sigma }_{k+1}^{z,\pm } 
+\hat{\sigma }_{k-1}^{z,\pm } -2\hat{\sigma }_k^{z,\pm } \to 
a^2\frac{\partial ^2\hat\sigma^{z,\pm } (z)}{\partial z^2}, 
\end{equation}
the 
equation, which, like to (\ref{eq14}), in compact vector form is:
\begin{equation}
\label{eq17}
\frac{\partial \hat {\vec {\sigma }}(z)}{\partial t}=\left[ {\hat {\vec 
{\sigma }}(z)\times \gamma _{_E} \vec {E}} \right]-\frac{2a^2J_{_E} }{\hbar 
}\left[ {\hat {\vec {\sigma }}(z)\times \nabla ^2\hat {\vec {\sigma }}(z)} 
\right],
\end{equation}
where 
\begin{equation}
\label{eq17a}
\begin{split}
\raisetag{60pt}
\vec {E} = E_{1} e^{i\omega t}\vec {e}_- + E_{1} e^{-i\omega t}\vec{e}_+ + \left( {\frac{-\omega _0 }{\gamma _{_E} }} \right)\vec {e}_z \\ \equiv E^+ \vec {e}_- + E^- \vec{e}_+ + E^z\vec {e}_z.
\end{split}
\end{equation}
The structure of vector $\vec {E}$ clarifies its physical meaning. Two components 
$E^+,\,\,E^-$ are right- and left-rotatory components of oscillating external 
electric field, third component $E^z$ is intracrystalline electric field, 
which produces two level energy splitting for each of the elementary unit of a chain 
system with value, equal to $\hbar \omega _0 $. When using  the analogy with magnetic resonance  the following 
relationship can be entered
\begin{equation}
\label{eq17b}
 \omega _0 =\frac{1}{\hbar }g_{_E} \beta _{_E} E_0 
=\gamma _{_E} E_0,
\end{equation}
 where $\beta _{_E} $ is electrical analogue of Bohr magneton, 
$g_{_E} $ is electrical analogue of magnetic $g$ -tensor, which is assumed for the 
simplicity to be isotropic. In other words, by means of relationship \ref{eq17b} the 
correspondence between an unknown intracrystalline electric field $E_0 $ and 
observed frequency $\omega _0 $ is set up.
Further, we take into consideration, that physical sense of operators $\hat 
{\vec {\sigma }}(z)$ in continuous limit is remained, that is, for each point 
of $z$ the components $\widehat{\sigma }^\pm (z)$, $\widehat{\sigma }^z(z)$ are 
satisfying to algebra, which is isomorphic to algebra of the set of spin 
components. Then by means of relationship, which has mathematically the same 
form for both the types of dynamical systems studied $\hat {\vec {S}}(z)\sim \frac{\hbar \hat {\vec {\sigma }}(z)}{2}$, the equation of the motion for operators of magnetic and electric spin moments are obtained. So, for example, the equation of the motion for electric spin moment operator is:
\begin{equation}
\label{eq18}
\frac{\partial \hat {\vec {S}}(z)}{\partial t}=\left[ {\hat {\vec 
{S}}(z)\times \gamma _{_E} \vec {E}} \right]-\frac{4a^2J_{_E}}{\hbar ^2}\left[ {\hat 
{\vec {S}}(z)\times \nabla ^2\hat {\vec {S}}(z)} \right]
\end{equation}

\section{L-L Equation Solution and Discussion}
We see, that transition operator method, being applying to optical transitions, leads unambiguously to conclusion on existence of electric spin moment, that is to the same conclusion, which was done by Dirac in 1928 in its famous paper \cite{Dirac}, although Dirac himself was  in some doubt about physical and even mathematical correctness of given conclusion. Transition operator method relieves all doubts in mathematical correctness of Dirac' conclusion (although method does not give any indications about imaginary or real character for given quantity).  Given aspect was discussed in recent paper \cite{Yearchuck_PL}, where experimental confirmation for electric spin moment discovery was represented. 
  
Equation (\ref{eq18}) gives for the case $J = 0$ quantum-mechanical optical analogue 
of classical Landau-Lifshitz equation in continuous limit. In fact equation(\ref{eq18}) and its magnetic analogue (if replace electrical characteristics by corresponding magnetic
characteristics by means of eq.(\ref{eq3})) are 
operator equations, which argue the correctness of eqs.(\ref{eq1}, \ref{eq2}), that is, the 
physical correctness of gyroscopic model for description of spectroscopic  
transitions and transitional 
phenomena for both optical and magnetic resonance spectroscopy.
 If $J \ne 0$ equation (\ref{eq18}) gives quantum-mechanical operator analogue of the
 equation for observables, which was introduced by Kittel for ferromagnetic spin wave resonance (FM SWR) description \cite{Kittel}. The equation (\ref{eq18}) (if put aside the 
operator symbols and to replace electrical characteristics by corresponding magnetic
characteristics) and the equation, introduced by Kittel, are coinciding 
mathematically to factor two in the second term. This difference is like to 
well known difference of gyromagnetic ratios for orbital and spin angular 
moments.  Given comparison means, that equation (\ref{eq18}) is the equation for optical analogue of ferromagnetic spin wave resonance. In other words, we have obtained the equation, which predicts new physical phenomenon - ferroelectric spin wave resonance (FE SWR). experimentally and reported in \cite{Yearchuck_Yerchak}. 
 
We can also conclude, that the spin value can be registered in solids directly by optical methods. Really, it will be possible in the cases, when the responce of dynamical system studied  can be obtained by registration schema, which takes into account the imaginary nature of electric spin moment. Given schema has to  ensure $\frac{\pi}{2}$  hyperbolic phase rotation in the functional space, which is isomorphic to Minkowski space like to microwave $\frac{\pi}{2}$ phase rotation by detection of the signals, being proportional to real and imaginary parts of complex magnetic susceptibiliy in electron spin resonance experiments.  On the other hand spin registration can be provided automatically, if in Hamiltonian there is item with product of the components of electric spin moment like to Hamiltonian (\ref{eq10}). Consequently,  transition operator method is theoretical substantiation for experimental methods of spin detection. Thus along with magnetic resonance methods we can detect a spin value of 
particles, quasiparticles, impurities or other centers in solids by optical 
methods, for example by study of transitional optical analogues of magnetic resonance and especially simply by new predicted phenomenon FE SWR. 

It should also be noted, that above considered theoretical 
description of FE SWR allow predict the difference in splitting constants 
which characterize FE SWR by its experimental detection with using of 
one-photon methods like to IR-absorption or IR-reflection and with using of 
two-photon methods like to Raman scattering. It is evident, that equations 
(\ref{eq17}), (\ref{eq18}) can immediately be used for single transition methods, for 
instance, for IR-absorption. By Raman scattering we have two subsequent 
transitions. Then operator $\hat {\vec {\sigma }}(z)$, which characterizes 
the transition dynamics by Raman scattering process, has to be consisting of
 two components $\hat {\vec {\sigma }}_1 (z)$ and $\hat {\vec {\sigma 
}}_2 (z)$ characterizing both the transitions, taken separately, that is 
$\hat {\vec {\sigma }}(z)=\hat {\vec {\sigma }}_1 (z)+\hat {\vec {\sigma 
}}_2 (z)$. Consequently, the equation for transition dynamics of second 
component, which will determine experimentally observed FE SWR-spectrum, is
\begin{gather}
\label{eq19}
\frac{{\partial \hat {\vec {\sigma}} _2 (z)}}{{\partial t}} = \left[ {\hat {\vec {\sigma}} _2 (z) \times \gamma _{_E} \vec E} \right] - \frac{{2a^2 J}}{\hbar}\left[ {\hat {\vec {\sigma}} _2 (z) \times \nabla ^2 \hat {\vec {\sigma}} _2 (z)} \right] \nonumber\\
 - \frac{{2a^2 J}}{\hbar }\left[ {\hat {\vec {\sigma}} _2 (z) \times \nabla ^2 \hat {\vec {\sigma}} _1 (z)} \right].
 \end{gather}
The second and the third items in (\ref{eq19})  are practically equal to each 
other, since we are dealing with interacting electric dipole moments of the 
same chain, that is, $\nabla ^2\hat {\vec {\sigma }}_1 (z)$ and $\nabla 
^2\hat {\vec {\sigma }}_2 (z)$ have almost equal values. Then we obtain, that 
the value of splitting constant by Raman scattering detection of FE SWR in the 
same sample is almost double in comparison with that one by IR detection of 
FE SWR. The observation of doubling in the splitting constant by Raman 
FE SWR-studies is additional direct argument in FE SWR identification.

 FE SWR was discovered experimentally and reported in \cite{Yearchuck_Yerchak}. 
  To compare with experiment the equation (\ref{eq10}) and consequently final equation (\ref{eq18}) have to be modified by taking into consideration 
the relaxation processes and from operator variables we have to proceed to observables. Modification was done in \cite{Yearchuck_Yerchak} fenomenologically by adding the term, which is similar to corresponding term in Bloch equations. 
If $J \ne 0$ equation (\ref{eq18}) gives quantum-mechanical operator analogue of the
 equation for observables, which was introduced by Kittel for ferromagnetic spin wave resonance (FM SWR) description \cite{Kittel}. It can be seen, that the coincidence of equation (\ref{eq18}) and Kittel equation \cite{Kittel} will take place, if put aside the 
operator symbols and to replace electrical characteristics by corresponding magnetic
characteristics. Consequently, to proceed to observables it is sufficient to put aside the 
operator symbols. It was suggested in \cite{Yearchuck_Yerchak}.
Then, for description of optical transition dynamics the equation 
will be 
\begin{equation}
\label{eq21}
\begin{split}
\raisetag{40pt}
\frac{\partial \vec {S}(z)}{\partial t} = \left[ {\vec {S}(z)\times \gamma_{E} 
\vec {E}} \right]\\ - \frac{4a^2J_{E}}{\hbar ^2}\left[ {\vec {S}(z)\times \nabla 
^2\vec {S}(z)} \right] + \frac{\vec {S}(z)-\vec {S}_0 (z)}{\tau },
\end{split}
\end{equation}
where  
 $\vec {S}_0 (z)$ is equilibrium value of electrical spin moment 
vector function, $\tau$ is relaxation time. 
It was also suggested in \cite{Yearchuck_Yerchak} that experinental data on FE SWR, which was detected by three various optical methods - by IR absorption, by IR reflection and by Raman scattering - will be well described by  linearized equation, corresponding to equation (\ref{eq18}).
 The linearized equation in \cite{Yearchuck_Yerchak} is considered. It was obtained under the assumption, that the values of oscillating external electric field 
components $E^x, E^y$ in $(\ref{eq21})$ are in experiment greatly less in 
comparison with the value of intracrystalline electric field component 
$E^z$, under analogous assumption relatively the components of total 
electrical spin moment and under additional assumption, that equilibrium distribution of $\vec {S}_0 (z)$ along the 
chain is homogeneous. All the assumptions are entirely correct for IR measurements. The linearized equation was solved  and 
 the relationships for a shape and 
amplitudes of resonance modes and dispersion law were obtained. They are
\begin{equation}
\label{eq22}
a_{n} = \left\{ {{\begin{array}{*{40}c}
 {-\frac{i \gamma_{E} S \tau^2 E_{1}}{\pi n} \frac{\left[{(\omega_n - \omega
) - \frac{i}{\tau}} \right]}{\left[ {1 + (\omega_{n} - \omega)^2 \tau^2} 
\right]},\,\,n = 1,\,3,\,5,... \hfill} \\
{\,0,\,\,\,\,\,\,\,\,\,\,\,n = 2,\,4,\,6,... \hfill}, \\
\end{array} }} \right. 
\end{equation}
\begin{equation}
\label{eq23}
\nu _n =\nu _0 - \mathfrak{A} n^2,
\end{equation}
where $n\in N$ including zero, $\nu_{n}$ is a frequency of \textit{n-th} mode, $\mathfrak{A}$ is 
a material parameter ($\mathfrak{A} = \frac{2 \pi {a}^{2} S \left|J\right|}{{\hbar L}^{2}} {n}^{2} > 0$).
Here $Re\,a_n $ is proportional to absorption signal, $Im\,a_n 
$ is proportional to dispersion signal.
It was found the followig. 1) Dispersion law (\ref{eq23}) is held true both by IR- and RS-detection of FE SWR. 2) The excitation of the only uneven modes by IR- detection (by which the experimental conditions were corresponding to applicability of linearized equation) takes place in accordance with (\ref{eq22}). 3) Inversely 
proportional dependence at resonance of the amplitudes of modes on mode number $n$ in accordance with (\ref{eq22}) is also held true, however also the only by
IR-detection of FE SWR. 4) Splitting of 
Raman active vibration modes is characterized by value of parameter $\mathfrak{A}$, being 
approximately by factor 2 greater, than parameter $\mathfrak{A}$, which characterizes IR FESWR spectra 
(by the frequencies of zero modes reduced by means of linear approximation 
procedure to the same value). The fourth result agrees well with the above discussed results, preliminary 
reported in \cite{Yearchuck_Doklady}. 
The agreement of theoretical and experimental results, which is rather well, allow to obtain new substantial information on carbynoid structure. Since Hamiltonian (\ref{eq10}) is true the only for the centers with the spin of $\frac{1}{2}$ it means that spin-Peierls solitons  \cite{Ertchak_Carbyne_and_Carbynoid_Structures},  \cite{Ertchak_J_Physics_Condensed_Matter} both optical and magnetical, which are formed in carbynoid chains have spin S = $\frac{1}{2}$. The observation of two types of FM SWR extitations \cite{Ertchak_J_Physics_Condensed_Matter} and two types of FE SWR extitations with slightly different splitting parameters $\mathfrak{A}^H$  and $\mathfrak{A}^E$ indicate unambiguously that  spin-Peierls solitons are produced separately in $\pi_x$= and $\pi_y$-subsystems of carbynoid electronic $\pi$-system. Moreover we can also conclude, that $\pi_x$- and $\pi_y$-subsystems of carbynoids are inequivalent.  It seems to be essential, that given inequivalency was observed after 12 Y storage. The only one set of  FM SWR has been found in the samples, which were studied immediately after their production (\cite{Ertchak_Carbyne_and_Carbynoid_Structures}, \cite{Ertchak_J_Physics_Condensed_Matter}), indicating on full equivalency of $\pi_x$- and $\pi_y$-subsystems of starting carbynoids. Naturally, all electronic properties, wich determined mainly by $\pi_x$- and $\pi_y$-subsystems are also changed with storage time. However, clear above presented elucidation of the cause for the changes of electronic properties gives key to production of storage stable carbynoids. The values of splitting parameters $\mathfrak{A}^E$ and $\mathfrak{A}^H$  allow to find 
 the ratio $J_{E }/J_{H}$ of exchange 
constants. The 
values of splitting parameter $\mathfrak{A}^H$ of FM SWR which was observed strictly in the same sample, that is on the same chains, were reported earlier in   \cite{Ertchak_J_Physics_Condensed_Matter}. The range of the ratio $J_{E }/J_{H}$ is $(1.2 - 1.6)10^{4}$. Given result seems to be direct proof,  that the function, which is invariant under gauge transformations is two component, that is complex-valued function. In other words, the complex charge corresponds 
to presence of exchange interacting solitons with two independent exchange constants. We can evaluate the ratio of imagine $e_{H} \equiv g$ to real $e_{E}\equiv e $  components of complex charge to be $\frac{g}{e} \sim \sqrt{J_{E }/J_{H}} \approx (1.1 - 1.3)10^{2}$. 
\begin{equation}
\label{eq24}
\begin{split}
\raisetag{40pt}
&\frac {\partial \sigma^{+}(z,t)} {\partial t} = i\omega_0\sigma^{+}(z,t) + i\Omega_R e^{i\omega t}\sigma^{z}(z,t) \\ +
& \frac{2ia^2J}{\hbar} \left[\sigma^{+} (z,t)\frac{\partial ^2\sigma^{z}(z,t)}{\partial z^2}-\sigma^{z}(z,t)\frac{\partial ^2\sigma^{+}(z,t)}{\partial z^2} \right],
\end{split}
\end{equation}
\begin{equation}
\label{eq25}
\begin{split}
\raisetag{40pt}
&\frac {\partial \sigma^{-}(z,t)} {\partial t} = -i\omega_0\sigma^{-}(z,t) -i\Omega_R e^{-i\omega t}\sigma^{z}(z,t)\\
& - \frac{2ia^2J}{\hbar} \left[\sigma^{-}(z,t) \frac{\partial ^2\sigma^{z}(z,t)}{\partial z^2}-\sigma^{z}(z,t)\frac{\partial ^2\sigma^{-}(z,t)}{\partial z^2} \right],
\end{split}
\end{equation}
\begin{equation}
\label{eq26}
\begin{split}
\raisetag{40pt}
&\frac {\partial \sigma^{z}(z,t)} {\partial t}  = 2i\Omega_R \left(\sigma^{+}(z,t) e^{-i\omega t} - \sigma^{-}(z,t) e^{i\omega t} \right) \\+ &\frac{4ia^2J}{\hbar} \left[ \sigma^{-}(z,t) \frac{\partial ^2\sigma^{+}(z,t)}{\partial z^2} - \sigma^{+}(z,t) \frac{\partial ^2\sigma^{-}(z,t)}{\partial z^2} \right].
\end{split}
\end{equation}
Let proceed to new variables 
\begin{equation}
\label{eq27}
\begin{split}
\raisetag{40pt}
&S^{+}(z,t) = \frac{\hbar}{2} \sigma^{+}(z,t) e^{-i\omega t},\\
&S^{-}(z,t) = \frac{\hbar}{2} \sigma^{-}(z,t) e^{i\omega t}, \\
&S'^{z}(z,t) = \frac{\hbar}{2} \sigma^{z}(z,t),
\end{split}
\end{equation}
then we will have
\begin{equation}
\label{eq28}
\begin{split}
\raisetag{40pt}
&\frac {\partial S^{+}(z,t)} {\partial t} = i\Delta  S^{+}(z,t) + i\Omega_R S'^{z}(z,t) \\
&+ \frac{4ia^2J}{\hbar^2} \left[ S^{+}(z,t) \frac{\partial ^2 S'^{z}(z,t)}{\partial z^2} - S'^{z}(z,t) \frac{\partial ^2 S^{+}(z,t)}{\partial z^2} \right],
\end{split}
\end{equation}
\begin{equation}
\label{eq28}
\begin{split}
\raisetag{40pt}
&\frac {\partial S^{-}(z,t)} {\partial t} = -i\Delta  S^{-}(z,t) - i\Omega_R S'^{z}(z,t) \\
&+ \frac{4ia^2J}{\hbar^2} \left[ S^{-}(z,t) \frac{\partial ^2 S'^{z}(z,t)}{\partial z^2} - S'^{z}(z,t) \frac{\partial ^2 S^{-}(z,t)}{\partial z^2} \right],
\end{split}
\end{equation}
\begin{equation}
\label{eq29}
\begin{split}
\raisetag{40pt}
&\frac {\partial S'^{z}(z,t)} {\partial t} = 2i\Omega_R \left[S^{+}(z,t) - S^{-}(z,t)\right]\\
&+\frac{4ia^2J}{\hbar} \left[S^{-}(z,t) \frac{\partial ^2 S^{+}(z,t)}{\partial z^2} - S^{+}(z,t) \frac{\partial ^2 S^{-}(z,t)}{\partial z^2}\right],
\end{split}
\end{equation}

where $\Delta = \omega_0 - \omega$. 

Passing on to new basis $\vec{e}_x$, $\vec{e}_y$, $\vec{e}_z$ and designating

\begin{equation}
\label{eq30}
\begin{split}
\raisetag{40pt}
&S_{x}(z,t) = \frac{2a^2J}{\hbar^2}(S^{+}(z,t) + S^{-}(z,t)),\\
&S_{y}(z,t) = i\frac{2a^2J}{\hbar^2}(S^{-}(z,t) - S^{+}(z,t)),\\
&S_{z}(z,t) = \frac{4a^2J}{\hbar^2}S'_{z}(z,t),
\end{split}
\end{equation}
we obtain 
\begin{equation}
\label{eq31}
\begin{split}
\raisetag{40pt}
&\frac {\partial S_{x}(z,t)}{\partial t} = -\Delta S_{y}(z,t)\\
&-\left[S_{y}(z,t) \frac{\partial^2 S_{z}(z,t)}{\partial z^2} - S_{z}(z,t) \frac{\partial ^2 S_{y}(z,t)}{\partial z^2}\right],
\end{split}
\end{equation}
\begin{equation}
\label{eq32}
\begin{split}
\raisetag{40pt}
&\frac {\partial S_{y}(z,t)}{\partial t} = \Delta S_{x}(z,t) + \Omega_R S_{z}(z,t) \\
&+ \left[ S_{x}(z,t) \frac{\partial^2 S_{z}(z,t)}{\partial z^2} - S_{z}(z,t)\frac{\partial^2 S_{x}(z,t)}{\partial z^2}\right],
\end{split}
\end{equation}
\begin{equation}
\label{eq33}
\begin{split}
\raisetag{40pt}
&\frac {\partial S_{z}(z,t)}{\partial t} = -\Omega_R S_{y}(z,t) \\
&+ \left[S_{y}(z,t)\frac{\partial^2 S_{x}(z,t)}{\partial z^2} - S_{x}(z,t)\frac{\partial^2 S_{y}(z,t)}{\partial z^2}S\right],
\end{split}
\end{equation}
or in vector form
\begin{equation}
\label{34}
\frac {\partial \vec{S}(z,t)}{\partial t} = \left[\vec{S}(z,t) \times \frac{\partial^2 \vec{S}(z,t)}{\partial z^2}\right] + \left[\vec{S}(z,t) \times \vec{\Omega}\right],
\end{equation}
\begin{equation}
\label{35}
\vec{S}(z,t) = S_{x}(z,t) \vec{e}_x + S_{y}(z,t) \vec{e}_y + S_{z}(z,t) \vec{e}_z,
\end{equation}
\begin{equation}
\label{36}
\vec{\Omega} = \Omega_R \vec{e}_x - \Delta \vec{e}_z.
\end{equation}
The equation (\ref{eq34}) has exact analytical n-soliton solution \cite{Takhtajan}. In particular, at  $n = 1$, the solution is
\begin{equation}
S_{x}(z,t) = \sin\theta \cos \varphi, S_{y}(z,t) = \sin\theta \sin \varphi, S_{z}(z,t)= \cos \theta, 
\end{equation}
where 
\begin{equation}
\label{eq37}
\begin{split}
\raisetag{40pt}
&\cos\theta = 1 - 2\beta^2 ch^{-2} [\beta\sqrt{\omega}(z-vt-z_0)],\\ 
&v = 4Re[\xi], \omega = 4|\xi|^2,
\end{split}
\end{equation}
\begin{equation}
\label{eq38}
\begin{split}
\raisetag{40pt}
&\varphi = \varphi_0 + \omega t + v (z-vt)/2 \\
&+ arctg \{\beta(1 - \beta^2)^{-1/2}th[\beta\sqrt{\omega}(z-vt-z_0)]\},
\end{split}
\end{equation}
\begin{equation}
\label{eq39}
\begin{split}
\raisetag{40pt}
&\beta = z_0 = ln\{|m|/(2Im[\xi])]\}/( 2Im[\xi]),\\
&Im[\xi]/|\xi|,   \varphi_0 = arg[m]. 
\end{split}
\end{equation}
Here $\xi$ is eigenvalue, $m$ is asymptotic characteristic of eigenfunction of differential operator  $L(z,t) = (\vec{S(z,t)} \vec{\sigma_P}) \partial_z$, $\vec{\sigma_P}$ is Pauli matrix vector.
The soliton is characterised by four real
parameters: velocity $v$, frequency $\omega$, phase $\varphi_0$ and
coordinate of the center $z_0$.

It should be noted, that cosideration of both variants of solutions, that is exact solution and solution of linearized equation, seems to be necessary. It follows from general virial theorem, which recently was proven by Buniy and Kephart.  For any classic field, independent of a partial model, the theorem  establishes, in particular, the lower energy bounds for massive classical solitons or lumps \cite{Buniy}. It means, that soliton solution will be applicable, when EM-field energy exceeds some threshold.

Let us remember that exact solution and solution of linearized equation were obtained by means of proceeding to observables in operator equation in s simple way - by change of operators into corresponding observables. Agreement with Kittel theory and, the essentials,  agreement with experimental data justify given suggestion, however mathematical proof is desirable. On second hand EM-field is relativistic object. It means, that the equations, describing the processes of interaction with EM-field, strictly speaking, have to be Lorentz-invariant. To solve given tasks we use further quaternion formalism. When applying to quantum mechanics quaternion formalism realizes its generalization, consisting in replacing of standard Hilbert space over the field of complex numbers into Hilbert space over the ring of quaternions. More strictly, biquaternions are used. We use the quaternion basis $\{e_i\}, i = \overline{0,3}$ with algebraic operations between elements, satisfying to relationships 
\begin{equation}
\label{eq40}
e_i e_j = \varepsilon_{ijk}e_k + \delta_{ij} e_{_0}, e_{_0} e_i = e_i, {e_{_0}}^2 = e_{_0}, i,j,k =\overline{1,3},
\end{equation}
where $\varepsilon_{ijk}$ is completely antisymmetric Levi-Chivita 3-tensor.

Then we use the equation (\ref{eq17}), completed like to (\ref{eq18}) by item, taking phenomenologically the relaxation processes 
\begin{equation}
\label{eq41}
\begin{split}
\raisetag{40pt}
&\frac{\partial\hat{\vec {\sigma}}(z,t) }{\partial t} = \left[\hat{\vec{ 
\sigma}}(z,t) \times \vec{F}(z,t) \hat{E} \right]\\
&- \frac{2a^2J_{_E} }{\hbar} \left[\hat{\vec{\sigma}}(z,t) \times \nabla ^2\hat{\vec {\sigma}}(z,t)\right] + \frac{{\vec{\sigma}}(z,t) - \vec{\sigma}_{_0}(z)}{\tau},
\end{split}
\end{equation}
where
$\vec{F} = \gamma _{_E} \vec {E}$, $\hat{E}$ is unit operator.  Quaternion algebra and algebra of Pauli matrix, completed in the manner above described, are isomorphic, see for example \cite{Berezin}. It means, that algebra of transions operators is also isomorphic to algebra of complete Pauli matrix set.
It is reasonable to introduce the quaternion operators by means of mappings
$\hat{E} \rightarrow \hat{e_{_0}},\hat{\vec{\sigma}}(z,t) \rightarrow i \hat{\vec{e}}, \hat{\vec{e}} = (\hat{e}_1,\hat{e}_2,\hat{e}_3)$ with conditions
$\forall{f(x)}$ the relationships $\hat{e}_k f(x) = {e}_k f_{k}(x)$  is taking place. Here $\Phi(x) = f_{_0}(x) e_{_0} + f_{i}(x) e_i$ is arbitrary quaternion function of quaternion argument  $x = x_{_0} + \vec{x}$, $x_{_0} = ict$, that is, components of $x_i \in ^{1}R_4$. 
Let us define now wave function $\Phi(x)$ of the chain system interactig with EM-field and described by Hamiltonian (\ref{eq10}) to be vector of the state in Hilbert space over quaternion ring. In other words, it is quaternion function of quaternion argument and can be represented in a form 

$ \Phi(x) = \Phi_{_0}(x) e_{_0} + Phi_{i}(x) e_i $. 

 It reazonable to suggest, that $\Phi_{_0}(x) = 0$. Then taking into account replacing 
$ t \rightarrow ict = x_{_0}$, equation (\ref{eq41}) can be rewritten in the form
\begin{equation}
\label{eq42}
\begin{split}
\raisetag{40pt}
&i c \frac{\partial(\hat{\vec{e}} |\Phi(x)|)}{\partial x_{_0}} = \left[\hat{\vec{e}} |\Phi(x)| \times \vec{F}(x) e_{_0}\right]\\
&- \frac{2a^2J_{_E} }{\hbar} \left[\hat{\vec{e}} |\Phi(x)| \times \nabla ^2(\hat{\vec{e}} |\Phi(x)|)\right] + \frac{\hat{\vec{e}} |\Phi(x)| - \hat{\vec{e}} |\Phi_{_0}(x)|}{\tau},
\end{split}
\end{equation}
where $|\Phi_{_0}(x)|$ is $|\Phi(x)|$ at $t = t_{_0}$, $\vec{F}(x)$ is vector quaternion delivered in conformity with vector $ \vec{F}(z,t)$.
It is is evident, that
$\vec{e} |\Phi(x)| = \vec{\Phi}(x)$. 
Then we have
\begin{equation}
\label{eq42}
\begin{split}
\raisetag{40pt}
&\frac{\partial\vec{\Phi}(x)}{\partial x_{_0}} = \frac{1}{ic} \left[\vec{\Phi}(x) \times \vec{F}(x)\right]\\
&- \frac{2  a^2 J_{_E}}{\hbar c} \left[\vec{\Phi}(x) \times \nabla ^2 \vec{\Phi}(x)\right] + \frac{\vec{\Phi}(x) -  \vec{\Phi}_{_0}(x)}{i c \tau},
\end{split}
\end{equation}
The equation obtained is Lorenz invariant equation for vector of the state $\vec{\Phi}(x)$ of the system EM-field + chain. Quaternion vector of the state $\vec{\Phi}(x)$ is transformed according to vector representation of Lorenz group, which is $\vec{\Phi'}(x) = L \vec{\Phi}(x) L*$. Spin vector $\vec{S}(x)$ is transformed under the same vector representation of Lorenz group. So it means, on the one hand, that (\ref{eq42}) will held true, if $\vec{\Phi}(x)$ replace into $\vec{S}(x)$, and we obtain therefore the equation for observables.
On the other hand, $\vec{S}(x)$ can be considered to be quaternion vector of the state of the system studied.

\section{Conclusions}

Therefore, quantum-mechanical analogue of Landau-Lifshitz equation has been derived with clear physical meaning of the quantities for both radio- and optical spectroscopy. It has been established that Landau-Lifshitz equation is fundamental physical equation underlying the dynamics of spectroscopic transitions and transitional phenomena. The solution of both linearized and starting nonlinear equation is presented. New phenomenon - ferroelectric spin wave resonance and its some spectroscopic peculiarities are described.  The agreement of theoretical and experimental results, which is rather well, allows to obtain new substantial information on chain structure and the properties of chain elementary units. For example, for the first time from pure optical measurements the value of spin $S = 1/2$ for optically active centers - spin-Peierls solitons has been obtained. The chains of  
spin-Peirls solitons are exchange coupled, the interaction is characterized by complex-valued charge function. The ratio of imagine  to real   components of complex charge is evaluated to be $\frac{g}{e} \approx (1.1 - 1.3)10^{2}$.

Lorenz invariant equation for vector of the state of the system EM-field + chain has been obtained and solved. It is shown, that spin vector can be considered to be quaternion vector of the state of the system studied. 
\begin{acknowledgments}
The authors are thankful to Professor, Corresponding Member of National Academy of Sciences of RB, L.Tomilchick for the helpful discussions of results.
\end{acknowledgments}

\end{document}